# Comment on slope of nonlinear FN plot and field enhancement factor

Weiliang Wang and Zhibing Li[*]

State Key Laboratory of Optoelectronic Materials and Technologies

School of Physics and Engineering, Sun Yat-sen University, 510275 Guangzhou, P.R. China
Email: stslzb@mail.sysu.edu.cn

Abstract:
It is common practice to extract field enhancement factor from the slope of FN plot. Many experimentalists working on field electron emission had reported multi-(linear segment) FN plots, which can be divided into several (usually two) linear segments. Then multi-(field enhancement factor) were extracted from the FN plot. They claimed that the field enhancement factor increases with applied field if the FN plot bends downward (vice versus if the FN plot bends upward). We show that this is contrary to fact.
Keywords: field emission, FN plot, field enhancement factor
PACS: 79.70.+q

## 1. Introduction

Many experimentalists working on field electron emission had reported multi-(linear segment) FN plots, which can be divided into several (usually two) linear segments [1-10]. Then multi-(field enhancement factor $\beta$) were extracted from the FN plot. They claimed that $\beta$ increases with applied field if the FN plot bends downward (vice versus if the FN plot bends upward). The present paper aims to illustrate that this is contrary to fact.

## 2. Field enhancement factor

According to the FN equation (For simplicity, we use the elementary FN-type equation. Experimentalists are recommended to use the technically complete FN-type equation [11]) which gives the local emission current density (LECD) $J_L$ in terms of the local work function $\phi$ and the applied macroscopic field $F_M$

$$J_L = a\phi^{-1}\beta^2 F_M^2 \exp\left[-b\phi^3/\beta F_M\right] \qquad (1)$$

where $a$ ($\cong 1.541434\,\mu A \cdot eV \cdot V^{-2}$) and $b$ ($\cong 6.830890\,eV^{-3/2} \cdot V \cdot nm^{-2}$) are universal constants, the field enhancement factor

$$\beta = -b\phi^3 / S_L \qquad (2)$$

where $S_L$ is the slope of an FN plot ($\ln J_L / F_M^2$ versus $1/F_M$). Some experimentalists applied this to each segment of a multi-(linear segment) FN plots. Therefore they found that $\beta$ is different in different $F_M$.

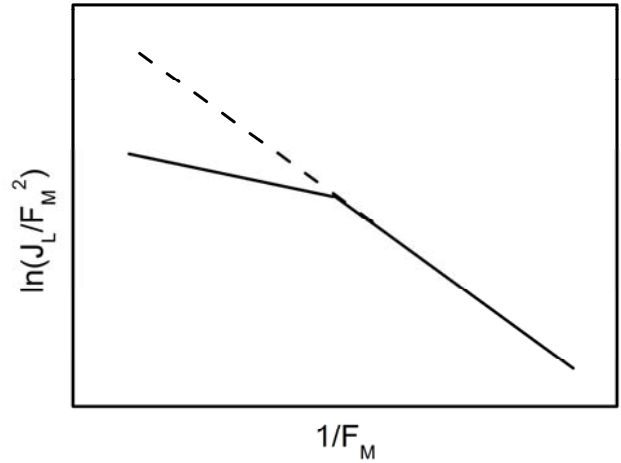

Figure 1. Illustration of a two linear segment FN plot.

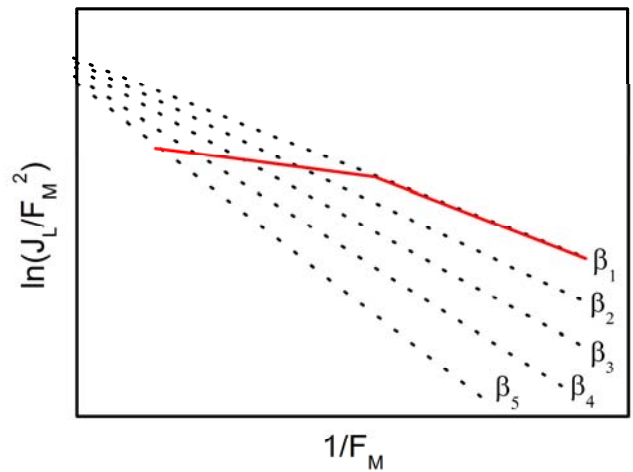

Figure 2. Illustration of a series of FN plots with field enhancement factor $\beta_1 > \beta_2 > \beta_3 > \beta_4 > \beta_5$ (dotted) and a bending downward FN plot (solid).

Let's take a two linear segment FN plot as an example (figure 1). This type of FN plots is reported in many experimental papers (for example [1, 4]). We call this bending downward FN plot. $|S_L|$ is greater (less) in low (high) field region. One would found that $\beta$ in high field region is greater than that in low field region. This is contrast to a simple reasoning: greater $\beta$ should lead to higher emission current, and then the FN plot in high field region should be above the dashed line (figure 1). The reason is that Eq. (2) is valid only if $\phi$ and $\beta$ are independent of $F_M$. Figure 2 illustrates intuitively how $\beta$ varies with $F_M$ in bending downward FN plots. The field enhancement factor is $\beta_1$ in low field region; therefore the FN plot (solid) coincides with the upper most dotted line. The field enhancement factor decreases to $\beta_2, \beta_3, \beta_4$ and $\beta_5$ successively in high field region, therefore the FN plot (solid) cross the lower dotted line corresponding to $\beta_2, \beta_3, \beta_4$ and $\beta_5$ in sequence. Therefore a bending downward (upward) FN plot means $\beta$ decrease (increase) with $F_M$ (if all other parameters are constant).

The present paper does not aim to discuses the reason of the variation of $\beta$. Because it had been extensively discussed in many literatures, the reasons might be resistance [12, 13], space charge effect [14], gas absorption [15], structure change of emission site [16, 17], non-uniformity of emission sites [18, 19], localized states [20], non-Schottky-Nordheim barrier [21, 22] or interaction between emitters [23].

## 3. Work function

The above discussion can be applied to work function $\phi$ straightforwardly. It is straightforward to see that a bending downward (upward) FN plot means $\phi$ increase (decrease) with $F_M$ (if all other parameters are constant).

The potential barrier can be lowered by the applied field [12, 24-26], thus $\phi$ may decrease with $F_M$. And $\phi$ can keep constant if the Fermi level is pinned in conduction band or large amount of surface states. What would the FN plot look like if $\phi = \phi_2$ in low field region and $\phi = \phi_1$ in high field region ($\phi_1 < \phi_2$)? It should coincide with the FN plot with $\phi = \phi_2$ in low field region and jump to the FN plot with $\phi = \phi_1$ in high field region. It would be a zigzag FN plot (figure 3). These zigzag FN plots were reported in many experimental papers (for example [27]). The present paper provides a possible explanation.

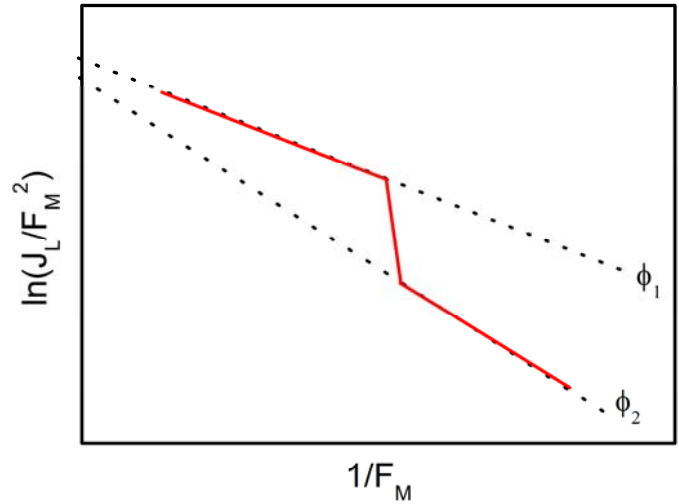

Figure 3. Illustration of a series of FN plots with local work function $\phi_1 < \phi_2$ (dotted) and a zigzag FN plot (solid).

## 4. Conclusion

It is inappropriate to extract field enhancement factor or work function from multi-(linear segment) FN plot. We show intuitively that a bending downward (upward) FN plot means $\beta$ decrease (increase) with $F_M$ (if all other parameters are constant), and a bending downward (upward) FN plot means $\phi$ increase (decrease) with $F_M$ (if all other parameters are constant). Zigzag FN plot may be due to a step function like work function (versus the applied field).

**Acknowledgments**

The project was supported by the National Basic Research Program of China (Grant No. 2007CB935500 and 2008AA03A314), the National Natural Science Foundation of China (Grant No. 11274393 and 11104358). The authors thank Prof. Juncong She for inspiring discussions.